\documentclass[twocolumn,showpacs,preprintnumbers,amsmath,amssymb,superscriptaddress]{revtex4}
\usepackage{bm}
\usepackage{graphicx}
\usepackage{dcolumn}
\usepackage{amsmath,amssymb,ams,bbm,epsfig}

\def\ed{\end{document}}

\def\bet{\begin{tabular}}
\def\eet{\end{tabular}}
\def\beqn{\begin{eqnarray}}
\def\eeqn{\end{eqnarray}}
\def\beq{\begin{equation}}
\def\eeq{\end{equation}}
\def\ba{\begin{array}}
\def\ea{\end{array}}
\def\bee{\begin{enumerate}}
\def\eee{\end{enumerate}}

\begin{document}
\preprint{PREPRINT ExB}
\title{Ericson fluctuations in an open, deterministic quantum system:\\
  theory meets experiment} 
\author{Javier Madro\~nero}
\affiliation{Max-Planck-Institut f\"ur Physik komplexer Systeme, Dresden}
\affiliation{Physik Department, Technische Universit\"at M\"unchen, 85747 Garching, Germany}
\author{Andreas Buchleitner}
\affiliation{Max-Planck-Institut f\"ur Physik komplexer Systeme, Dresden}
\date{\today}
\begin{abstract}
We provide numerically exact photoexcitation cross sections of rubidium
Rydberg states in crossed, static electric and magnetic fields, in
quantitative agreement with recent experimental results. 
Their spectral
backbone underpins a clear transition towards
the Ericson regime. 
\end{abstract}
\maketitle

The experimental characterization and theoretical understanding of complex
quantum transport phenomena is of fundamental relevance for many, otherwise
rather remote research areas which exploit quantum interference effects 
for the purpose of an ever improving control over the quantum dynamics of
increasingly complicated systems
\cite{qc97,abu98b,mueller02,akab03,wimberger03,hornberger03,wellens04,paul05}.  In such
context, 
``complexity'' may arise 
from many particle interactions, from deterministic chaos, or from disorder,
to name just a few of its possible causes.  
Notwithstanding the many origins of complex quantum dynamics, its qualitative 
macroscopic signatures are often very similar: an
enhanced sensitivity on small changes in some control parameter (be it the
boundary conditions in a disordered mesoscopic conductor \cite{pichard90}, 
the perturbation
strength in a strongly driven atomic system \cite{abu98b}, the injection
energy in a 
compound nucleus reaction \cite{bohigas91}, or the observation angle in a
random 
laser \cite{cao99,viviescas03}) enforces a 
statistical description such as to
isolate robust quantities to characterize the underlying physical
processes. Surprisingly, many of the resulting predictions are universal in
character, i.e., they apply to, on a first glance, rather different classes of
physical
objects, which only share an increased density of states, and the
nonperturbative coupling of their various degrees of freedom.         

While erratic fluctuations of some experimental observable 
under changes of a control parameter come not too
surprising in many-particle dynamics or in disordered systems
\cite{akab03,qc97,borgonovi92,pichard90}, they still 
remain rather counterintuitive and for many a cause of discomfort 
in simple quantum systems with only few degrees
of freedom -- think of single electron or photon transport across  two dimensional billiards
\cite{sachrajda98,schanze01,dembowski05}, 
or of the ionization probability of a one electron Rydberg state under
external forcing \cite{casati90,abu98b}. Here, classically chaotic dynamics
substitute 
for disorder 
and many-particle interactions, though are expected to generate very similar
-- if not the same -- statistical behaviour, in tantalizing contrast, e.g.,
to the 
clock-like regularity of Kepler like Rydberg motion. Hitherto, however, 
experimental evidence for chaos-induced fluctuations in the coherent quantum
transport 
in low dimensional, strictly deterministic systems is scarce
\cite{sachrajda98}, since bona fide 
transport measurements require very high spectral resolution in the continuum
part of the spectrum, together with the continuous tunability of a suitable
control 
parameter. 

In the present Letter, we focus on a paradigmatic example in the realm of
atomic physics -- the photoexcitation of one electron Rydberg states in the
presence of crossed, static electric and magnetic fields. Our contribution is
motivated by recent experimental results \cite{stania05,stania05b} 
which probe the atomic spectrum above
the field induced ionization saddle, and refines the interpretation of
the experiments as the first observation of Ericson fluctuations in a strictly
deterministic, open quantum system. Furthermore, this represents the first
full-fledged, parameter-free quantum treatment of the truly three dimensional
crossed fields problem at experimentally realistic spectral densities.

Ericson fluctuations are a
universal 
statistical feature of strongly coupled, fragmenting quantum systems, first
predicted \cite{ericson60} and observed \cite{brentano64} 
in compound nuclear reactions. They manifest as
fluctuations in the excitation cross sections into the regime of highly
excited, metastable resonance states, with typical decay widths larger than
the average level spacing, such that single maxima in the cross section cannot
be identified with single resonances any more, but are rather due to the
interference of several of them. 
In particular, this implies that the typical scale of fluctuations induced by
interfering decay channels is {\em smaller} than the typical width 
of individual resonances.
In quantum systems with a well-defined
classical analog, Ericson fluctuations can be understood as a hallmark of
chaotic scattering \cite{bluemel88}, somewhat complementary to the
exponential 
or algebraic 
decay of the survival probability on the time axis
\cite{abu95b}.  
A similarly paradigmatic case of the
Ericson scenario as the one to be considered presently 
arises in the photoexcitation of highly doubly excited states of helium, yet
still awaits its full experimental confirmation, due to the extraordinary
requirements on the experimental resolution \cite{puettner01}.

Let us start with the Hamiltonian describing the single electron Rydberg
dynamics  subject to
crossed electric and magnetic fields, in atomic units, and assuming an
infinite mass of the nucleus:
\begin{equation}
H=\frac{{\bf p}^2}{2}+V_{\rm atom}(r)+\frac{1}{2}B\ell_z+\frac{1}{8}B^2(x^2+y^2)+Fx\, . 
\label{ham}
\end{equation}
Here, $B$ and $F$ are the strength of magnetic and electric field,
respectively, and $\ell_z$ the angular momentum projection on the magnetic
field 
axis. Note that no uniquely defined one particle potential is available for
the bare atomic Hamiltonian with its alkali earth multielectron core, and
$V_{\rm 
  atom}$ is therefore not given explicitely. 
However, the deviation of $V_{\rm atom}$ from a strictly Coulombic potential 
in a small but finite volume
around the nucleus can be accounted for by the phase shift experienced by the
Rydberg electron  upon scattering off the multielectron core
\cite{halley92,krug02}. This phase shift 
is fixed by the $\ell$-dependent quantum defects $\delta_{\ell}$ of the
unperturbed atom, which are precisely determined by spectroscopic data
\cite{lorenzen83}.  
The continuum
coupling induced by the electric field (which lowers the ionization threshold
by creating a Stark saddle in the $x$ direction) is incorporated into our
theoretical treatment by complex dilation of the Hamiltonian, and the
numerical diagonalization of the resulting complex symmetric  
matrix (represented in
a real Sturmian basis) 
immediately yields the full spectral structure (eigenenergies $E_j$, decay
rates 
$\Gamma_j$,
and eigenvectors $|E_j\rangle $) underlying
the experimentally measured photoionization cross section, without adjustable
parameters \cite{halley92,krug02}. Note that, due to the mixing of all good
quantum numbers of the 
unperturbed dynamics (the principal
quantum number $n$, the angular momentum quantum number $L$, and the angular
momentum projection $M$) by the crossed external fields, 
we are dealing with a
truly three dimensional problem \cite{main94,milscewski96}, and the only
remaining 
constants of the 
motion are the energy $E$, and parity
with respect to the plane defined by the
magnetic field axis. Consequently, the complex symmetric eigenvalue problem
which we have to solve has a respectable size, with typical (sparse banded) 
matrix dimensions
of $300000\times 4000$, for the energies and field strengths we shall be
dealing 
with hereafter.

The experiments in \cite{stania05,stania05b} probe the energy range from
$-57.08\ {\rm cm}^{-1}$ to $-55.92\ {\rm cm}^{-1}$ (corresponding to
principal quantum numbers $n\simeq 43\ldots 45$ of the bare atom), at fixed
electric 
field strength $F=22.4\ \rm kV/m$, and for three different values of the
magnetic field, $B=0.9974\ \rm T$, $B=1.49\ \rm T$, and $B=2.0045\ \rm T$. The
electric field shifts the effective ionization threshold to $-91.4\ {\rm
  cm}^{-1}$, hence the experimentally probed energy range lies clearly in the
continuum part of the spectrum. Invoking the
scale invariance of the classical dynamics in a Coulomb potential, 
the
specific 
choice
of $F$ and $B=2.0045\ \rm T$
is equivalent to the one
in
\cite{main94} (there for the purely Coulombic problem, and for a lower lying
energy range, $n\simeq 19\ldots 22$, i.e., at much reduced spectral densities), and 
corresponds to 
classically chaotic scattering (where electric and
magnetic field are of comparable strength, though incompatible symmetry).
While the finite size multielectron core of rubidium
strictly speaking invalidates such a scaling argument (as well as it blurres a
strict quantum-classical analogy, simply due to the absence of a well defined
classical one particle analog) \cite{krug02}, it may still serve as
an approximate guide into the regime of broken symmetries of the quantum
problem \cite{delande94b}. 

We performed numerical diagonalizations of the complex dilated Hamiltonian
(\ref{ham}) {\em precisely} for the experimental parameter values, though in a
broader energy range, such as to
illustrate the emergence of Ericson fluctuations from a smooth continuum
background, with 
increasing Rydberg energies. The photoexcitation cross section $\sigma(E)$ is
readily obtained from the quantum spectrum, via
\begin{equation}
\sigma(E)=\frac{4\pi (E-E_0)}{c\hbar}{\rm
  Im}\sum_j\frac{D_{j;L=2}^2}{E_j-i\Gamma_j/2-E}\, , 
\label{cross}
\end{equation}
where $D_{j;L=2}$ denotes the relative oscillator strength \cite{halley92} of
the transition 
from the initial state $|n=5\ L=1\ M=-1\rangle $ with energy $E_0\sim -0.002$\,a.u. 
into the electronic eigenstate
$|E_j\rangle$ with decay rate $\Gamma_j$, mediated by a single photon linearly
polarized along the magnetic field axis (thus selecting the odd parity part
of the spectrum). 
Note that our computational method
does not 
allow for an 
absolute calibration of the oscillator strengths, since the wave function of
$|n=5\ L=1\ M=-1\rangle $ is not explicitely known. For the technical details
underlying the expression (\ref{cross}), we refer the reader to
\cite{halley92,abu94}. 

\begin{figure}[t]
\epsfig{file=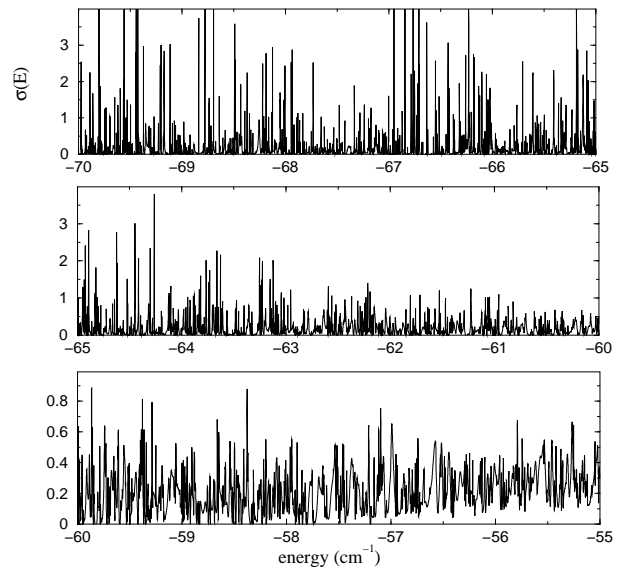,angle=-90,width=0.45\textwidth}
\caption{Numerical photoexcitation cross sections, at magnetic and electric
  field strengths $B=2.0045\ \rm T$ and $F=22.4\ \rm kV/m$, respectively, in
  the energy range $E=-70.0\ldots -65.0\ {\rm cm}^{-1}$ (top), $E=-65.0\ldots
  -60.0\ {\rm cm}^{-1}$ (middle), and $E=-60.0\ldots -55.0\ {\rm cm}^{-1}$
  (bottom). The latter completely covers the energy range probed in the
  experiments reported in \protect\cite{stania05,stania05b}. While individual
  resonances are well resolved in the lower lying spectra, on top of a flat
  continuum background, a strongly fluctuating continuum structure in the
  bottom plot expresses the increasing contribution of overlapping
resonances on the
  spectral level, see Fig.~\ref{fig2}.
\label{fig1}}
\end{figure}
Figure~\ref{fig1} shows the thus obtained photoexcitation spectra, at magnetic
and 
electric field
strengths $B=2.0045\ \rm T$ and $F=22.4\ \rm kV/m$, respectively, and in three
different energy ranges, $-70.0\ldots 
-65.0\ {\rm cm} ^{-1}$, $-65.0\ldots -60.0\ {\rm cm} ^{-1}$, and $-60.0\ldots -55.0\
{\rm cm} ^{-1}$. 
The latter of these completely covers the experimentally probed
energy 
interval. Clearly, individual resonances can be resolved in the two lower
lying spectra, on top of an essentially flat continuum background. In
contrast, the experimentally probed energy range is characterized by a
strongly fluctuating continuum, with only few narrow structures on top, what
immediately suggests the overlapping of an appreciable part of the resonances which
contribute to the sum in eq.~(\ref{cross}). Inspection of the
underlying
distribution of resonance widths $\Gamma_j$ along the energy
axis, measured in units of the average local level spacing $\Delta =\langle
E_j-E_{j-1}\rangle$, indeed comforts this picture, see Fig.~\ref{fig2}: 
\begin{figure}[t]
\epsfig{file=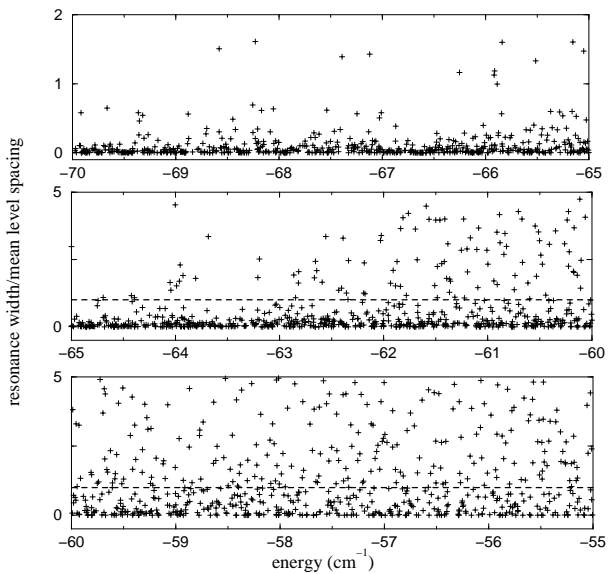,angle=-90,width=0.45\textwidth}
\caption{Distribution of the resonance widths $\Gamma_j$ contributing to the
  photoexcitation cross section, eq.~(\ref{cross}), in units of the local mean
  level spacing $\Delta =\langle E_j-E_{j-1}\rangle$, over an energy range which spans
  the domain probed in the three plots of fig.~\ref{fig1}, for the same values
  of $F$ and $B$. The dashed line at 
$\Gamma_j/\Delta =1$ separates isolated ($\Gamma_j/\Delta <1$) 
from overlapping resonances ($\Gamma_j/\Delta >1$). Comparison with
  Fig.~\ref{fig1} shows how broad resonances on the spectral
  level induce a strongly fluctuating continuum
  background, in the photoexcitation cross section. The energy range
  $E=-60.0\ldots -55.0\ {\rm cm}^{-1}$, where resonance overlap is most
  pronounced in this plot, covers the energy range probed in the experiments
  reported in \cite{stania05,stania05b}.
\label{fig2}}
\end{figure}
The weight of large resonance widths with $\Gamma_j>\Delta$ clearly
increases as we probe higher lying energies, and amounts to approx. $65\%$
of all contributing resonances, in the experimentally probed energy
range. Many of the structures in $\sigma(E)$ are consequently due to the
interference of decay channels through overlapping resonances.

A close comparison of 
our numerical cross section in the lower panel of Fig.~\ref{fig1} (and of
Fig.~\ref{fig3} below) shows close similarity with the experimental signal
\cite{stania05}, 
though no perfect coincidence is achieved. 
\begin{figure}[t]
\epsfig{file=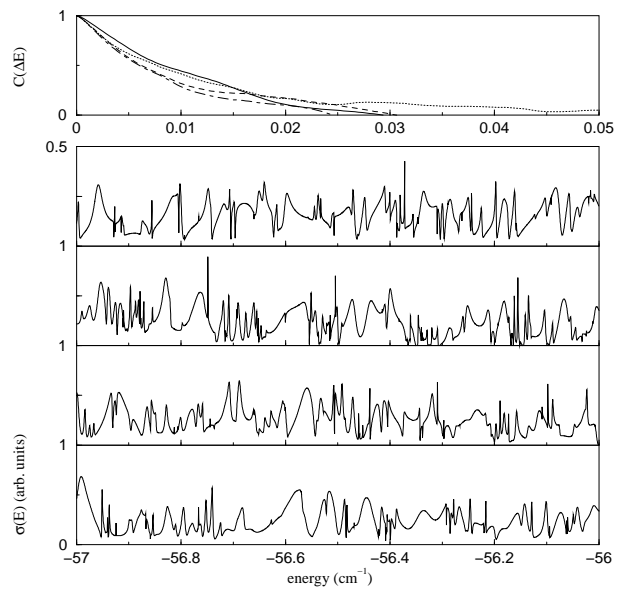,angle=-90,width=0.45\textwidth}
\caption{Top: (Normalized) Autocorrelation functions $C(\Delta E)$ of the photoionization cross sections for
  magnetic field strengths $B=2.0045\ \rm T$ (full line), $B=1.49\ \rm T$
  (dashed line), $B=0.9974\ \rm T$ (dotted line), and $B=0.563\ \rm T$
  (dash-dotted line), at electric field
  strength $F=22.4\ \rm kV/m$. The corresponding cross sections are shown in
  the subsequent panels, with increasing magnetic field strength from top to
  bottom.  
The characteristic correlation lengths $\gamma$ of $C(E)$ deduced from
  our -- parameter-free -- numerical treatment, for the three
  largest values of $B$, are in 
very good agreement with the experimentally reported values, see
  table \ref{table}. Note that even for the smallest value of $B$ (not
  studied in \cite{stania05}) there is a largely fluctuating continuum
  background, as opposed to the purely Coulombic problem realized with
  hydrogen atoms \cite{main94}.
\label{fig3}}
\end{figure}
This, 
however, is anything but surprising, precisely due to the characteristic, 
extreme sensitivity
of quantum spectra and cross sections with respect to tiny changes in the boundary
conditions, in the regime of classically chaotic
dynamics \cite{abu98b,bohigas91,casati90,sachrajda98}. 
Therefore, rather than scanning parameter space on a fine mesh, to reproduce
the experimentally observed (but fragile!) cross section exactly
\cite{stevens96}, we 
calculate the 
autocorrelation function $C(\Delta E)=\langle (\sigma(E+\Delta
E)-\langle\sigma\rangle)(\sigma(E)-\langle\sigma\rangle\rangle$ of
$\sigma(E)$, which is predicted \cite{bluemel88} 
to have a
Lorentzian shape with the characteristic width $\gamma$, $C(\Delta
E)\sim 1/(\Delta E^2+\gamma^2)$, in the regime of stronlgy overlapping
resonances. 
The latter condition is indeed met by all the three values of the
magnetic field employed in the experiments in \cite{stania05,stania05b}, and  
Fig.~\ref{fig3} shows the autocorrelation functions,  
together with the
excitation spectra from which they are deduced.
$\gamma$ is expected to be a
statistically robust quantity, and we verified that its value remains 
unaffected by the sensitive
parameter dependence of $\sigma(E)$ itself, within the error bars indicated. 
The respective values of
the characteristic widths $\gamma$ are in perfect agreement with the
experimental values, as listed in table I.
\begin{table}
\begin{tabular}[t]{|l||l|l|l|}
\hline\hline $B\ \rm [T]$ & $\gamma^{\rm exp}\ \rm [cm^{-1}]$ \protect\cite{stania05} &
$\gamma^{\rm th}\ \rm [cm^{-1}]$ \\\hline \hline
$0.9974$ & $0.0083$ & $0.0082\pm 0.0005$  \\\hline
$1.49$ & $0.0065$ & $0.0062\pm 0.0005 $  \\\hline
$2.0045$ & $0.0081$ & $0.0080\pm 0.0005$  \\\hline \hline
\end{tabular}
\caption{Comparison of the characteristic correlation decay length
  $\gamma^{\rm th}$ deduced from our numerical photoexcitation spectra
  with the experimental values $\gamma^{\rm exp} $ reported in
  \protect\cite{stania05}. 
  Within the indicated error bars, which are estimated from changes of
  $\gamma$ under small changes 
of $B$ and of the electric field strength $F$ (within their experimental
  uncertainties \cite{stania05}), the agreement is perfect. In particular,
  also the nonmonotonous dependence of $\gamma$ on the magnetic field
  strength $B$ is recovered. 
  \label{table}
}
\end{table}
In particular, also the nonmonotonous dependence of $\gamma$ on the magnetic field
  strength $B$ is recovered. 

To complete the picture, we also display in
  Fig.~\ref{fig3} the
  excitation spectrum and the associated 
  cross section at a weak magnetic field $B=0.563\ \rm T$ (not recorded in
  \cite{stania05}), 
where the classical
  dynamics of the associated Coulombic problem is near regular, since the
  electric field dominates the dynamics \cite{main94}. In contrast, the
  present result for rubidium exhibits a very similar structure as for
  stronger magnetic fields, certainly due to the destruction of the Coulomb
  symmetry by the multielectron core \cite{krug04}.

In conclusion, we revealed the spectral backbone of experimentally observed 
fluctuations in
the photoexcitation probability of nonhydrogenic rubidium Rydberg states in
crossed static electric and magnetic fields -- a truly three dimensional,
paradigmatic case of microscopic chaotic (half) scattering. By correlating the
experimentally available data with the resonance spectrum of the atom in the
field (obtained from an accurate numerical treatment without adjustable
parameters), 
and with the evolution of the latter along the energy axis, we
theoretically/numerically 
prove
that these experiments indeed successfully entered  
the regime of Ericson fluctuations, for the first time in a perfectly
deterministic, open quantum system.

We acknowledge support by the Rechenzentrum Garching of the Max Planck
Society, through access to the IBM Regatta system. 

\bibliography{literatur05}

\end{document}